\documentclass[galaxies,review,accept,oneauthor,pdftex,10pt,a4paper]{mdpi}

\usepackage{booktabs}
\usepackage{multirow}
\usepackage{soul}
\usepackage{microtype}

\usepackage{upgreek}
\usepackage{textcomp}

\firstpage{1}
\pubvolume{6}
\issuenum{4}
\articlenumber{138}
\pubyear{2018}
\copyrightyear{2018}
\history{Received: 14 September 2018; Accepted: 5 December 2018; Published: 9 December 2018}
\updates{yes}

\usepackage[flushleft]{threeparttable}
\usepackage{booktabs}

\DeclareUnicodeCharacter{FEFF}{}

\newcommand{\RNum}[1]{\uppercase\expandafter{\romannumeral #1\relax}}
\newcommand\ion[2]{#1$\;${%
\ifx\@currsize\normalsize\small \else
\ifx\@currsize\small\footnotesize \else
\ifx\@currsize\footnotesize\scriptsize \else
\ifx\@currsize\scriptsize\tiny \else
\ifx\@currsize\large\normalsize \else
\ifx\@currsize\Large\large
\fi\fi\fi\fi\fi\fi
\rmfamily\RNum{#2}}\relax}%

\newcommand\arcsec{\mbox{$^{\prime\prime}$}}%
\newcommand{\hst}{\ifmmode HST\else {\it HST}\fi}
\newcommand{\kms}{\ifmmode \mathrm{km~s^{-1}}\else km~s$^{-1}$\fi}
\newcommand{\lir}{\ifmmode L_\mathrm{IR}\else $L_\mathrm{IR}$\fi}
\newcommand{\lsun}{\ifmmode L_\odot\else $L_\odot$\fi}
\newcommand{\ha}{\ifmmode {\mathrm H}\alpha\else H$\alpha$\fi}
\newcommand{\lya}{\ifmmode {\mathrm Ly}\alpha\else Ly$\alpha$\fi}
\newcommand{\mdot}{\ifmmode \dot{M}\else $\dot{M}$\fi}
\newcommand{\mstar}{\ifmmode M_*\else $M_*$\fi}
\newcommand{\msun}{\ifmmode M_\odot\else $M_\odot$\fi}
\newcommand{\nt}{\ifmmode $[\ion{N}{2}]$\else[\ion{N}{2}]\fi}
\newcommand{\ntl}{\ifmmode
  $[\ion{N}{2}]~6583~\AA$\else[\ion{N}{2}]~6583~\AA\fi}
\newcommand{\ntll}{\ifmmode
  $[\ion{N}{2}]~6548,~6583~\AA$\else[\ion{N}{2}]~6548,~6583~\AA\fi}
\newcommand{\oth}{\ifmmode $[\ion{O}{3}]$\else[\ion{O}{3}]\fi}
\newcommand{\othl}{\ifmmode
  $[\ion{O}{3}]~5007~\AA$\else[\ion{O}{3}]~5007~\AA\fi}
\newcommand{\othll}{\ifmmode
  $[\ion{O}{3}]~4959,~5007~\AA$\else[\ion{O}{3}]~4959,~5007~\AA\fi}
\newcommand{\sfr}{\ifmmode \mathrm{SFR}\else SFR\fi}
\newcommand{\ssfr}{\ifmmode \mathrm{sSFR}\else sSFR\fi}
\newcommand{\sigsfr}{\ifmmode \Sigma_\mathrm{SFR}\else
  $\Sigma_\mathrm{SFR}$\fi} \newcommand{\smpy}{\ifmmode
  M_\odot~\mathrm{yr}^{-1}\else M$_\odot$~yr$^{-1}$\fi}
\newcommand{\vc}{\ifmmode v_\mathrm{circ}\else $v_\mathrm{circ}$\fi}
\newcommand{\vmax}{\ifmmode v_\mathrm{max}\else $v_\mathrm{max}$\fi}
\newcommand{\vout}{\ifmmode v_\mathrm{out}\else $v_\mathrm{out}$\fi}
\newcommand{\vfifty}{\ifmmode v_{50\%}\else $v_{50\%}$\fi}
\newcommand{\vninety}{\ifmmode v_{90\%}\else $v_{90\%}$\fi}
\newcommand{\vtwosig}{\ifmmode v_{98\%}\else $v_{98\%}$\fi}

\newcommand\aj{\it{Astron. J.}}
\newcommand\araa{\it{Annu. Revs. Astron.  Astrophys.}}
\newcommand\apj{\it{Astrophys. J.}}
\newcommand\apss{\it{Astrophys. Space
    Sci.}}
\newcommand\aap{\it{Astron. Astrophys.}}
\newcommand\mnras{\it{Mon. Not. R. Astron.
    Soc.}}
\newcommand\nat{\rm{\emph{Nature}}}
\newcommand\natas{\rm{\emph{Nat. Astron.}}}

\Title{A Review of Recent Observations of Galactic Winds Driven by
  Star Formation}

\Author{David S. N. Rupke\orcidA{}}

\AuthorNames{David S. N. Rupke}

\address[1]{Department of Physics, Rhodes College, Memphis, TN 38112,
  USA; drupke@gmail.com}

\abstract{Galaxy-scale outflows of gas, or galactic winds (GWs),
  driven by energy from star formation are a pivotal mechanism for
  regulation of star formation in the current model of galaxy
  evolution. Observations of this phenomenon have proliferated through
  the wide application of old techniques on large samples of galaxies,
  the development of new methods, and advances in telescopes and
  instrumentation. I review the diverse portfolio of direct
  observations of stellar GWs since 2010. Maturing measurements of the
  ionized and neutral gas properties of nearby winds have been joined
  by exciting new probes of molecular gas and dust. Low-$z$ techniques
  have been newly applied in large numbers at high $z$. The~explosion
  of optical and near-infrared 3D imaging spectroscopy has revealed
  the complex, multiphase structure of nearby GWs. These observations
  point to stellar GWs being a common feature of rapidly star-forming
  galaxies throughout at least the second half of cosmic history, and
  suggest that scaling relationships between outflow and galaxy
  properties persist over this period. The~simple model of a
  modest-velocity, biconical flow of multiphase gas and dust
  perpendicular to galaxy disks continues to be a robust descriptor of
  these flows.}

\keyword{galactic winds; feedback; outflows; star formation}

\begin{document}


\section{Introduction}

Significant amounts of gas in galaxies move in outward radial
trajectories due to energy imparted by star formation. This energy
originates from a combination of phenomena rooted in stellar
processes: radiation, winds, explosive events, and cosmic rays. These
gas outflows powered by star formation, or stellar galactic winds
(GWs), continue to be a dynamic topic of observational research in
the era of large galaxy surveys and multi-messenger
astronomy. Outflows are challenging to characterize largely because of
two factors: the large contrast between the outflow and an underlying
galaxy disk; and the complex, multiphase structure of outflows. High
quality data in many gas tracers are essential to adequately quantify
the mass, momentum, and energy budget of the wind.

Advances in observations of GWs driven by stellar processes in this
decade have come from the use of new observational techniques and the
application of old techniques to much larger samples. Both have been
aided by new-and-improved telescopes or instrumentation. In the first
category fall molecular gas transitions newly applied to study
outflows (notably the hydroxyl molecule), mid- and far-infrared
(MIR/FIR) imaging of dust, and resonant-line emission in the
rest-frame ultraviolet (UV). In the second category are surveys with
integral field spectrographs (IFSs), with multi-object long-slit
spectrographs, and with the Cosmic Origins Spectrograph (COS) on the
{\it Hubble Space Telescope (\hst)}.

Excellent and thorough reviews from previous decades of theory and
observations of GWs provide in-depth discussions of the quantities of
observational interest and the astrophysics of GWs
(e.g.,~\citep{1993ASSL..188..455H,2005ARAA..43..769V}). They also
show the trajectory and progress of the field. The~scope of this
review is narrower. I synthesize observational results from the
current decade (approximately 2010 through the present). I~focus on
direct measures of outflows and do not discuss studies that infer the
presence or properties of GWs by studying other phenomena.

As an example of an indirect measurement of the presence or properties
of GWs, studies of the mass-metallicity relation constrain how
efficiently GWs eject metals from galaxy disks compared to metal
production and reaccretion (e.g.,~\citep{2014ApSS.349..873Z}). A
second example is the significant reservoirs of highly ionized carbon
and oxygen that preferentially arise in the circumgalactic media of
actively star-forming galaxies
\citep{2011Sci...334..948T,2013ApJ...768...18B,2014ApJ...796..136B,2016ApJ...833..259B,2017ApJ...846..151H}. A
logical source of these metals is stellar GWs. Third, the hot halos
that appear to be a common feature of star-forming galaxies
\citep{2013MNRAS.435.3071L} may be produced by stellar GWs. Finally,
cosmological simulations typically use numerical prescriptions for the
unresolved physics of stellar GWs and compare simulated galaxy
properties with observed galaxy properties (like the galaxy mass
function) to constrain the nature of GWs
(e.g.,~\citep{2010MNRAS.406.2325O,2014MNRAS.445..581H,2014ApJ...782...84A}). Such
indirect measurements are essential for a complete picture of the
relationship of GWs to their surrounding environments. However,
at~present, interpretations of these measurements based on stellar GWs
typically compete with other physical models and are rarely
definitive.

Stellar processes are not the only possible drivers of GWs. Evidence
continues to accumulate that actively accreting supermassive black
holes (active galactic nuclei, or AGN) are also important in powering
GWs in galaxies with above-average masses (see reviews in
\citep{2012ARAA..50..455F,2017NatAs...1E.165H}). However, it can be
difficult to distinguish whether the AGN is energetically
important for the GW if significant star formation is also present,
except in the most powerful AGN. This difficulty is due to the
possible power sources being cospatial at low resolution and to the
uncertain duty cycle of AGN. Consequently, in this review I focus on
studies of purely star-forming systems, or at least those where star
formation clearly dominates the galaxy's luminosity. AGN with
low-to-moderate Eddington ratios are almost certainly present in many
galaxies classified as purely star-forming
\citep{2012ApJ...746...90A,2016ApJ...826...12J}, but their energetic
importance for GWs remains unquantified.

I organized this review at the highest level by redshift. This is
useful for two reasons. First, star formation and galaxy properties at
redshifts above unity differ significantly from those in the local
universe. The~global star formation rate (\sfr) peaked at $z\sim1.9$
(e.g.,~\citep{2014ARAA..52..415M}) and gas mass fraction increases
with increasing redshift
(e.g.,~\citep{2013ApJ...768...74T}). Galaxies may grow from the inside
out, meaning that disks are more extended compared to stars at high
$z$~\citep{2015ApJ...813...23V,2016ApJ...828...27N}. Second, low-$z$
winds are much better characterized because many more techniques can
be brought to bear. Some techniques are in practice easier to apply to
high-redshift galaxies, but on balance this is not the case.

\section{Winds Driven by Star Formation at Low redshift}

Most data on nearby stellar GWs in the first 20 years of earnest work
came from the optical and X-ray bands. These studies fell largely into
two classes: (1) long-slit spectroscopic and/or narrowband imaging
surveys
(e.g.,~\citep{1996ApJ...462..651L,1998ApJ...506..222M,2005ApJS..160..115R})
or (2) case studies using optical 3D spectroscopy, X-ray imaging
spectroscopy, or narrowband imaging
(e.g.,~\citep{1994ApJ...433...48V,1999ApJ...523..575L,2002ApJ...565L..63V,2009ApJ...697.2030S}). This
work was primarily focused on starburst galaxies, which lie above the
star-forming main sequence in \sfr. High-mass starbursts are typically
luminous in the infrared (IR) and classified as either luminous or
ultraluminous infrared galaxies. These LIRGs and ULIRGs are defined to
have $\lir>10^{11}$~\lsun\ and $\lir>10^{12}$~\lsun, which corresponds
to $\sfr>10$~\smpy\ and $\sfr>100$~\smpy\ if all the IR luminosity
is powered by star formation. Low-mass (dwarf) starbursts are luminous
in the UV and optical bands.

New observing capabilities, surveys, and archival databases have
allowed probes of more physical conditions, broadening the picture of
winds sketched first by small studies of the warm and hot ionized and
cool neutral phases. They have also extended the study of winds to
main-sequence galaxies.

This section is divided into four parts. First, I discuss new UV
surveys of nearby starbursts. Second, I visit the use of the Sloan
Digital Sky Survey (SDSS) for outflow studies. Third, I summarize
results from the widening use of integral field spectroscopy. I end by
surveying significant advances in revealing and quantifying the
molecular gas and dust in GWs using ground-based submillimeter and
space-based FIR telescopes.

\subsection{Ultraviolet Surveys} \label{sec:uv}

The rest-frame UV contains many interstellar absorption lines and
\lya, the brightest UV emission line in unobscured
starbursts. Previous UV instruments have been used to study individual
systems or small samples, but the sensitivity of COS has enabled
population studies of low-$z$, UV-bright starbursts at high
signal-to-noise (S/N). These studies are large enough to study
correlations between the properties of GWs and galaxy properties.

Two studies find a correlation between the velocity of the ionized
outflow and basic properties of the galaxy and starburst itself:
stellar mass \mstar\ and star formation rate \sfr\
\citep{2015ApJ...811..149C,2016ApJ...822....9H}. Table
\ref{tab:uvfits} compares these correlations and the ionization states
probed. The first \citep{2015ApJ...811..149C} uses COS data on 48
starbursts at $z < 0.26$ taken from the literature. The~second
\citep{2016ApJ...822....9H} combines COS and {\it Far-Ultraviolet
  Spectroscopic Explorer (FUSE)} data on 39 starbursts at $z < 0.25$
\citep{2015ApJ...809..147H} plus MMT or Keck spectra of 9 compact
starbursts at $z = 0.4-0.7$
\citep{2014MNRAS.441.3417S,2014Natur.516...68G}. Notably, the studies
differ in their definition of outflow velocity. The first
\citep{2015ApJ...811..149C} uses the 50\%\ and 90\%\ points in the
cumulative velocity distribution (CVD; starting from the red side of
the line). The second \citep{2016ApJ...822....9H} uses either the
98\%\ point of the CVD (in the case of data taken from secondary
sources \citep{2014MNRAS.441.3417S,2014Natur.516...68G}) or the
authors' own velocity measurements ($v_\mathrm{max}$). I label the
correlations from this second study with~$v_\mathrm{max}$ for
simplicity.

\begin{table}[H]
\centering
    \caption{Log-Log Fits of Outflow vs. Galaxy Properties. \label{tab:uvfits}}
\scalebox{0.95}[0.95]{\begin{tabular}{cccccccc}
      \toprule
      \textbf{Axes}	  & \textbf{Tracer} & \textbf{IP (eV)} & \textbf{$N$}& \textbf{Range} & \textbf{Slope} & \textbf{$p$} & \textbf{Reference} \\
      \midrule
      $\vfifty$ vs. \sfr & \ion{Si}{2} & 16.3 & 48 & $10^{-1}-10^2$ \smpy & 0.22 $\pm$ 0.04 & $<$0.001 &~\citep{2015ApJ...811..149C} \\
      $\vfifty$ vs. \sfr & \ion{Na}{1} & 5.1 & 41 & $10^{-1}-10^3$ \smpy & 0.35 $\pm$ 0.06 & $\cdots$ &~\citep{2005ApJ...621..227M} \\
      $\vfifty$ vs. \sfr & \ion{Na}{1} & 5.1 & 13 & $10^{0.8}-10^{2.2}$ \smpy & 0.15 $\pm$ 0.06 $^{c}$ & $\cdots$ &~\citep{2016AA...590A.125C} \\
      $\vfifty$ vs. \sfr & \ion{Na}{1} & 5.1 & 13 & $10^{0.8}-10^{2.2}$ \smpy & 0.30 $\pm$ 0.05 $^{c}$ & $\cdots$ &~\citep{2016AA...590A.125C} \\
      $\vninety$ vs. \sfr  & \ion{Si}{2} & 16.3 & 48 & $10^{-1}-10^2$ \smpy & 0.08 $\pm$ 0.02 & 0.002 &~\citep{2015ApJ...811..149C} \\
      $\vninety$ vs. \sfr  & \ion{Na}{1} & 5.1 & 59 & $10^{-1}-10^3$ \smpy & 0.21 $\pm$ 0.04 & $<$0.001 &~\citep{2005ApJS..160..115R} \\
      $\vninety$ vs. \sfr  & \ion{H}{1}, \ion{N}{2} & 13.6--29.6 & 48 & $10^{0.7}-10^{2.6}$ \smpy & 0.24 $\pm$ 0.05 & $<$0.001 &~\citep{2014AA...568A..14A} \\
      $\vmax$ vs. \sfr &   \ion{Si}{2}, \ion{C}{2}, \ion{Mg}{2} & 15.0--24.4 & 48 & $10^{-2}-10^3$ \smpy & 0.32 $\pm$ 0.02 & $<$0.0001 &~\citep{2016ApJ...822....9H} \\
      $\vfifty$ vs. \mstar & \ion{Si}{2} & 16.3 & 48 & $10^{9}-10^{11.5}$ \msun & 0.20 $\pm$ 0.05 & 0.002 &~\citep{2015ApJ...811..149C} \\
      $\vninety$ vs. \mstar & \ion{Si}{2} & 16.3 & 48 & $10^{9}-10^{11.5}$ \msun & 0.12 $\pm$ 0.03 & 0.003 &~\citep{2015ApJ...811..149C} \\
      $\vninety$ vs. \mstar $^b$ & \ion{Na}{1} & 5.1 & 52 & $\cdots$ & 0.28 $\pm$ 0.08 & $<$0.001 &~\citep{2005ApJS..160..115R} \\
      $\vfifty$ vs. \vc $^a$ & \ion{Si}{2} & 16.3 & 48 & $10^{1.8}-10^{2.5}$ \kms & 0.87 $\pm$ 0.17 & 0.002 &~\citep{2015ApJ...811..149C} \\
      $\vninety$ vs. \vc   & \ion{Na}{1} & 5.1 & 20 & $10^{1.4}-10^{2.7}$ \kms & 0.85 $\pm$ 0.15 & $<$0.001 &~\citep{2005ApJS..160..115R} \\
      $\vninety$ vs. \vc $^a$ & \ion{Si}{2} & 16.3 & 48 & $10^{1.8}-10^{2.5}$ \kms & 0.44 $\pm$ 0.09 & 0.003 &~\citep{2015ApJ...811..149C} \\
      $\vmax$ vs. \vc $^a$ & \ion{Si}{2}, \ion{C}{2}, \ion{Mg}{2} & 15.0--24.4 & 48 & $10^{1.3}-10^{2.5}$ \kms & 1.16 $\pm$ 0.37 & $<$0.0001 &~\citep{2016ApJ...822....9H} \\

      $\eta$ vs. \mstar & \ion{O}{1}, \ion{Si}{2}--\ion{Si}{4} & 13.6--45.1 & 7 & $10^7-10^{11}$~\msun & $-$0.43 $\pm$ 0.07 & $<$0.001 &~\citep{2017MNRAS.469.4831C} \\
      $\eta$ vs. \mstar & \ion{H}{1}, \ion{N}{2} & 13.6--29.6 & 33 & $10^{9.6}-10^{11.2}$~\msun & $-$0.43 & $\cdots$ & ~\citep{2014AA...568A..14A} \\
      $\eta$ vs. \mstar $^b$ & \ion{Na}{1} & 5.1 & 42 & $10^{10}-10^{11}$~\msun & $-$0.95 $\pm$ 0.20 & 0.006 & ~\citep{2005ApJS..160..115R} \\
      $\eta$ vs. \vc $^a$ & \ion{O}{1}, \ion{Si}{2}--\ion{Si}{4} & 13.6--45.1 & 7 & $\cdots$ & $-$1.56 $\pm$ 0.25 & $<$0.001 &~\citep{2017MNRAS.469.4831C} \\
      \bottomrule
    \end{tabular}}\\
   \begin{tabular}{cccccccc}
\multicolumn{1}{p{\textwidth -.88in}}{\footnotesize
      IP gives the ionization potential range of the atomic
      tracers; $N$ is the number of galaxies in the fit; the range
      applies to the galaxy property in the fit; the slope is for a
      linear fit in log-log space (or the exponent of a power law fit
      in linear space); and the $p$-value is the estimated likelihood
      of a null correlation. For references where only the correlation
      coefficient is provided, the $p$-value is inferred from it.
    $^a$ \vc\ is calculated from \mstar\ using a linear
      scaling.~\citep{2015ApJ...811..149C,2017MNRAS.469.4831C} use
      $\mathrm{log}~(\vc/\kms) = 0.28~\mathrm{log}(\mstar/\msun) -
      0.67$ from~\citep{2011MNRAS.417.2347R}, while
     ~\citep{2016ApJ...822....9H} uses
      $\mathrm{log}~(\vc/\kms) = 0.29~\mathrm{log}(\mstar/\msun) -
      0.79$.
    $^b$ \mstar\ is assumed to be proportional to the $K$-band
      luminosity.
    $^c$ The first of these fits uses integrated spectra; the
      second is from spatially resolved fits.}
    \end{tabular}
\end{table}

While both studies find correlations, the ranges of galaxy properties
in one are wider~\citep{2016ApJ...822....9H}. This study also finds
steeper slopes in lines fitted to the data; the slopes from the two
studies differ at $>1\sigma$. The~second study does not specify a
fitting method~\citep{2016ApJ...822....9H} , while the first uses a
method that accounts for $x$ and $y$ errors, outliers, and scatter
\citep{2015ApJ...811..149C}. Neither study finds a significant
correlation between outflow velocity and specific star formation rate
sSFR $\equiv \sfr/\mstar$. One also finds no correlation with star
formation rate surface density \sigsfr, but their sample has a range
$\sigsfr = 10^{-2}-10^{1}$
\smpy~kpc$^{-2}$~\citep{2015ApJ...811..149C}. The~second study adds
three orders of magnitude to the upper end of this range and do find a
correlation, which they parameterize as
$v_\mathrm{max} = 3296 /[ (\sigsfr /1307.9)^{-0.34} +
(\sigsfr/1307.9)^{0.15} ]$ at a significance of $p<0.0001$
\citep{2016ApJ...822....9H}.

The mass outflow rate \mdot\ and mass outflow rate normalized to the
star formation rate ($\eta\equiv\mdot/\sfr$)\footnote{This quantity
  has an uncertain physical interpretation, though at face value it
  might quantify the capability of a wind to act as negative feedback
  on star formation. The~logic is: if $\eta>1$, then more gas is
  leaving the region than is forming stars. Thus, the outflow is going
  to deplete gas more quickly than star formation (maybe leading to
  fewer stars in the end). It is sometimes referred to as the
  mass-loading efficiency or factor, perhaps suggesting that it
  measures the amount of gas that is ``loaded'' into the wind as it
  moves through the galaxy. That is, it is the ratio of the mass of
  gas that eventually emerges from star-forming regions to the mass of
  outflowing gas that is initially produced by star formation through,
  e.g., stellar winds and supernovae. However, it is not a direct
  measure of this. The~term mass-loading originated with studies of
  how much cool, ambient gas is mixed into a hot wind phase in the
  model of a wind driven by a hot fluid.} were computed for subsamples
of these two larger samples
\citep{2015ApJ...809..147H,2017MNRAS.469.4831C}. One study estimates
the wind column density using multiple ions and a stacked spectrum;
they apply this single column density to the entire sample of 39
galaxies~\citep{2015ApJ...809..147H}. Their mass outflow rate is then
a product of the outflow velocity and starburst radius times a
constant. They find a correlation between \mdot\ and \sfr\ that is
near-linear (though the slope is unquantified) and which appears to be
driven largely by the correlation between \mdot\ and
$v_\mathrm{max}$. They also find that $\eta$ is inversely correlated
with \sfr\ and \mstar\ (or \vc).

\begin{figure}[H]
\centering
\includegraphics[scale=0.8]{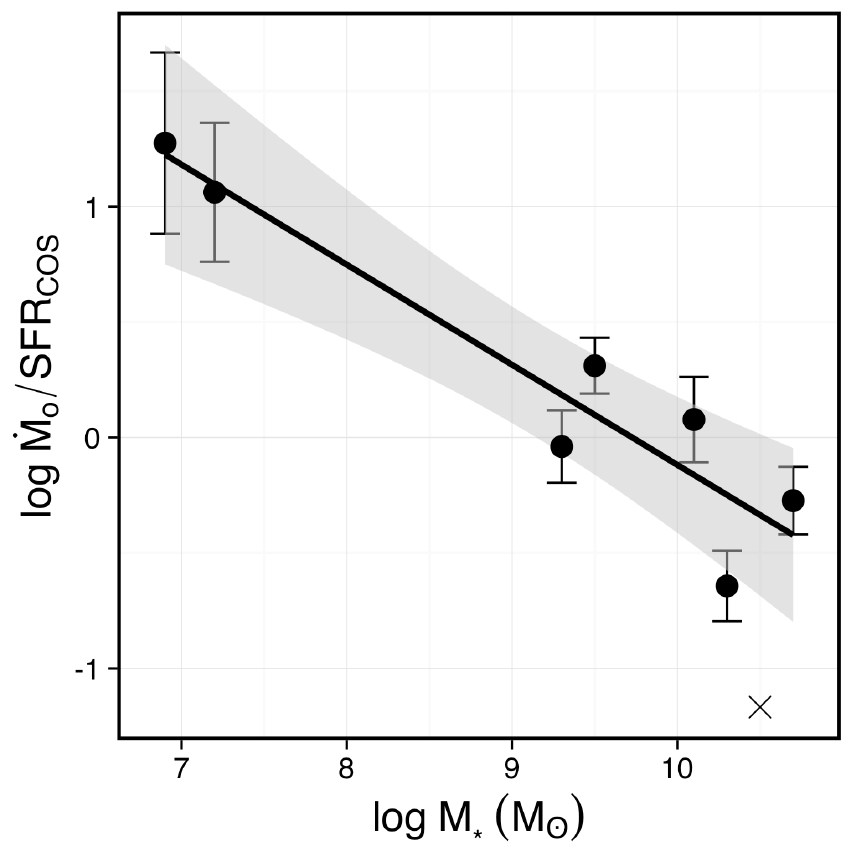}
\caption{A fit to $\eta\equiv\mdot/\sfr$ vs. \mstar\ for a sample of
  nearby GWs~\citep{2017MNRAS.469.4831C}. The~gray region is the 95\%\
  confidence interval. Outflow masses have been estimated from
  multiple rest-frame UV absorption lines and photoionization
  modeling. Relationships between outflow and galaxy properties are
  key tools for connecting observations to theory. The~fitted slope of
  $-$0.43 $\pm$ 0.07 compares favorably with predictions from
  numerical simulations at low masses
  ~\citep{2012MNRAS.421.3522H,2013MNRAS.436.1787L,2015MNRAS.454.2691M,2015MNRAS.452.1184M}. At
  higher masses the slope may be steeper, as~suggested by some
  simulations~\citep{2013MNRAS.436.1787L,2015MNRAS.452.1184M} and
  measurements~\citep{2005ApJS..160..115R}.  Reproduced with
  permission from Figure 1 of reference \cite{2017MNRAS.469.4831C}.}
\label{fig:eta}
\end{figure}

The other study finds a similar result for the inverse correlation of
$\eta$ and galaxy mass (\citep{2017MNRAS.469.4831C}; Figure~\ref{fig:eta}; Table \ref{tab:uvfits}). While based on a much smaller
sample size, this work uses high signal-to-noise spectra,
velocity-resolved optical depths and covering factors, and
photoionization modeling to estimate the ionization state, density,
metallicity, and inner radius of the wind (assuming a model relating
velocity and radius). The~measured slope ($-$0.43 $\pm$ 0.07) matches
predictions from numerical simulations, which are in the range $-$0.35
to $-$0.50 at low masses
\citep{2012MNRAS.421.3522H,2013MNRAS.436.1787L,2015MNRAS.454.2691M,2015MNRAS.452.1184M}.

The velocity measurements in these UV surveys employ lines from ions
of relatively low ionization potential
\citep{2015ApJ...811..149C,2016ApJ...822....9H}, while the estimated
wind masses are inferred from a larger range of ions
\citep{2015ApJ...809..147H,2017MNRAS.469.4831C}. The~resulting fits to
GW vs. galaxy properties are in overall agreement with similar fits
made a decade earlier using the cool, neutral gas phase (Table
\ref{tab:uvfits};
\citep{2005ApJS..160..115R,2005ApJ...621..227M}). The~spread in fitted
slopes illustrates the systematic uncertainties in this enterprise,
but a conservative synthesis points to velocity depending on \sfr\ and
\mstar\ as $v\sim\sfr^{0.1-0.3}$ and $v\sim\mstar^{0.1-0.3}$. The
dependence on circular velocity is thus steeper (close to linear). The
upper end of the range of slopes of $v$ vs. \sfr\ matches a prediction
from simulated absorption lines in 3D hydrodynamic simulations
\citep{2017ApJ...843..137T}. There are fewer measurements of how
$\eta$ depends on various properties, and the systematic uncertainties
in $\eta$ are higher than in velocity. However, as discussed above,
published fits are surprisingly consistent with predictions from
numerical simulations \citep{2015MNRAS.452.1184M,2015MNRAS.454.2691M},
including at high masses where the slope steepens
\citep{2005ApJS..160..115R,2015MNRAS.452.1184M}.

This begs the question of what drives the scatter in these
relationships. It is worth noting that the pursuit of scaling
relations relies on reducing the velocity field of a galactic outflow
to a single parameter in a single tracer. Outflows are unlike galaxies
in that they are inherently a violent, non-equilibrium process whose
structural properties are governed by time-varying power sources, gas
hydrodynamics, and radiation transfer rather than gravitational
processes. Their properties thus almost certainly depend on multiple
host parameters simultaneously. The~fact that outflow properties scale
at all with bulk galaxy properties such as \sfr\ is not surprising (more
energy means higher-velocity gas). There are clearly important details
beyond this, however, that are harder to quantify and may reflect the
unique structural properties or history of particular galaxies (Is it
a merger? Did the outflow just turn on?) or other hidden scalings that
are orthogonal to those with \sfr\ and mass. One solution might be to
combine multiple galaxy or wind properties into single parameters in
search of an ``outflow fundamental plane,'' but combining, e.g., two
parameters into one has yet to noticeably improve matters
\citep{2016ApJ...822....9H,2017MNRAS.469.4831C}.

The imprint of GWs in the UV is found not only absorption lines but
also in Ly emission lines and the escape of ionizing radiation beyond
the Lyman limit. Extended \lya\ emission and Lyman continuum (LyC)
escape may be caused in part by outflows that create low-density holes
in a galaxy's~ISM.

Recent studies of \lya\ and absorption lines in samples of order tens
of galaxies have quantified the impact of outflows on the properties
of \lya\ and the escape of LyC. Population studies of \lya\ in nearby
galaxies are, however, more complicated to interpret than studies of
absorption lines because of the resonant emission and absorption
behavior of \lya\ and its sensitivity to dust. As a result, these
studies have not reached firm conclusions on how these tracers reflect
GW properties. There is some suggestion that outflow velocity is
correlated with Ly$\alpha$ escape fraction
\citep{2013ApJ...765..118W,2015ApJ...803....6M,2015ApJ...805...14R,
  2015ApJ...809...19H,2015ApJ...810..104A,2017AA...605A..67C,2017ApJ...851L...9J}. However,
this correlation has significant scatter, and cases exist of low
outflow velocity and high LyC escape
\citep{2017ApJ...851L...9J}. \lya\ and outflow velocity are also both
correlated with \sfr\ and/or \sigsfr
~\citep{2017AA...605A..67C,2015ApJ...810..104A}. In one sample,
galaxies with and without escaping LyC have similar outflow velocities
\citep{2017AA...605A..67C}. The~simple presence of outflows may be a
necessary but not a sufficient condition for \lya\ and/or LyC escape
\citep{2015ApJ...805...14R}; other factors such as the outflow
acceleration, low \ion{H}{1} column density, or low metallicity may
also be required for \lya\ or LyC to exit the dense regions of a
galaxy
\citep{2015ApJ...803....6M,2015ApJ...809...19H,2017AA...605A..67C}.

\subsection{Single-Aperture Surveys} \label{sec:sdss}

Searches for outflows in single-aperture optical spectroscopic surveys
like SDSS are hampered by the spatially unresolved bright host
galaxy. Outflow features in emission lines are typically overwhelmed
by ionized gas emission from the star-forming host disk, which in
galaxies with modest GWs lies in the same velocity range as the
outflow. Interstellar absorption lines can be similarly dominated by
stellar absorption features.

Both limitations can be overcome with data of high enough S/N and
high-quality modeling of the underlying stellar continuum. Stacking of
many spectra is commonly used to achieve the required S/N, and stellar
models that match the spectral resolution of SDSS are mature. (Even if
these models may not uniquely constrain the star formation history or
stellar population properties, they fit the data very well.)

Three studies of GWs in star-forming galaxies have stacked SDSS DR7
data. The~authors of the first of these stack 10$^5$ high-mass
($\mstar\sim10^{10}-10^{11}$~\msun) galaxies in bins of various
physical parameters to study the properties of cool, neutral outflows
using the interstellar \ion{Na}{1}~D doublet
\citep{2010AJ....140..445C}. They detect this resonant line in both
absorption and emission. They find that the velocity and equivalent
width of absorption scale with inclination, such that face-on galaxies
are observed to have faster, higher equivalent width outflows,
consistent with the model of minor-axis outflows. Outflow equivalent
width and linewidth also correlate with \sigsfr\ (over the range
$10^{-2.5}-10^{-0.5}$~\smpy~kpc$^{-2}$) and \mstar\ (over the small
range $10^{10.3}-10^{11.2}$~\msun). However, systemic interstellar
absorption also scales with \sigsfr\ and \mstar, so the nature of
these correlations is uncertain.

The second study uses 200,000 galaxies to stack by \sfr\ and \mstar\
over a range of galaxy properties comparable to the ranges in samples
of individual galaxies (\citep{2016AA...588A..41C}; Table
\ref{tab:uvfits}; Sections \ref{sec:uv} and \ref{sec:ifs}):
$\sfr = 10^{-2.7}-10^{2.3}$~\smpy\ and
$\mstar = 10^{7.3}-10^{11.8}$~\msun. They fit \othl, H$\alpha$, and
\ntll\ using a high S/N instrumental profile and extract the
line-of-sight velocity distribution (LOSVD) of both the ionized gas
and stars. They use excess blueshifted emission-line gas at the
extreme end of the gas LOSVD to measure outflow velocity, with the
stellar LOSVD serving as a reference. For star-forming galaxies
selected by line ratio~\citep{2003MNRAS.346.1055K}, these authors find
that \vout\ correlates significantly with \sfr\ and \ssfr. \vout\ does
not correlate with \mstar, though it does with stellar velocity
dispersion, suggesting a discrepancy in how these quantities are
measured. This method is most sensitive to higher-velocity outflows,
and outflows are detected primarily at $\sfr>1$~\smpy\ and
$\ssfr > 10^{-9}$~yr$^{-1}$. They are also preferentially observed in
galaxies with \sfr\ values that put them above the main sequence.

Finally, a third study employs a similar technique over a larger
sample of 600,000 galaxies in bins of \sfr\ and \mstar\
\citep{2017AA...606A..36C}. These authors apply two methods for
parameterizing outflows: (1) they fit the \oth\ emission line in each
stack using two Gaussians; and (2) they use the observed \oth\ profile
as the gas LOSVD and calculate measures of line width and
asymmetry. They then remove the instrumental resolution in
quadrature. They find no significant detection of starburst-driven
winds in star-forming galaxies. However, they select star-forming
galaxies using a more restrictive line-ratio criterion
\citep{2006MNRAS.371..972S} than the ionized gas study that does
detect stellar GWs~\citep{2016AA...588A..41C}. They argue that a more
permissive selection~\citep{2003MNRAS.346.1055K} allows low-luminosity
AGN to contaminate a sample of star-forming galaxies; these AGN could
in turn produce outflows in galaxies that are supposedly purely
star-forming. Low-luminosity AGN exist throughout the star-forming
sequence with a range of contributions to \oth\
\citep{2012ApJ...746...90A,2016ApJ...826...12J}. It is certainly
plausible that star formation, rather than low-luminosity AGN, powers
the outflows detected in galaxies whose line emission is dominated by
star formation~\citep{2016AA...588A..41C}, but separating potential
contributions from low-luminosity AGN requires further work.

These stacking studies add weight to the conclusions drawn from
studies of smaller samples and individual galaxies by probing more
statistically complete samples of galaxies across a wider range of
basic galaxy properties.  There is agreement between two of the
stacking studies discussed above that ionized and/or neutral winds are
a common feature of galaxies of modest to high \sfr\ across the mass
spectrum~\citep{2010AJ....140..445C,2016AA...588A..41C}, consistent
with results discussed elsewhere in this review. The~disagreement with
the third study~\citep{2017AA...606A..36C} on this point is puzzling,
though the difficulty of separating AGN from star formation as a
possible energy source in galaxies with supermassive black holes
accreting at lower Eddington ratios is a valid concern. Furthermore,
the correlations between wind and galaxy properties are akin to those
seen using UV and other optical studies (Section \ref{sec:uv}; Table
\ref{tab:uvfits}), and the connection to inclination angle is
consistent with a wide range of observational results.

\subsection{IFS Results} \label{sec:ifs}

\begin{figure}[H]
  \centering
  \includegraphics[scale=0.45]{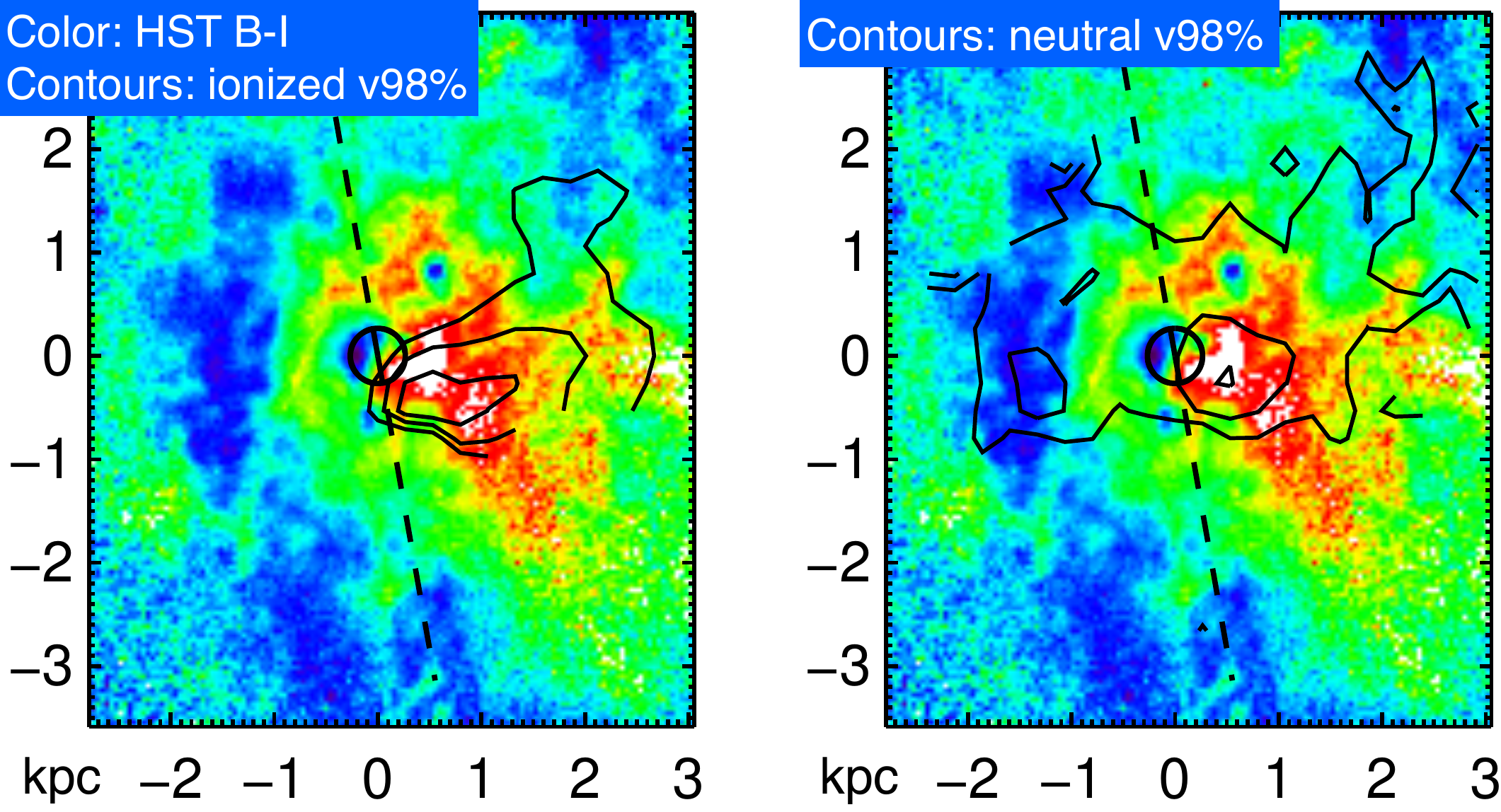}
  \caption{Illustration of the complex, multiphase structure of a
    nearby starburst-driven GW as observed with IFS (F10565$+$2448;
   ~\citep{2013ApJ...768...75R}). The~background color maps are \hst\
    images of $F435W-F814W$ (or~$B-I$), where red and white indicate the
    reddest colors. They show dusty filaments emerging from the
    obscured starburst along the minor axis in red and white; the near
    side of the star-forming disk is shown as the blue clumpy regions
    to the left in each panel. The~galaxy major axis is the dashed
    line, and the contours show ionized gas \vtwosig\ on the left
    ($-$800, $-$700, $-$600~\kms) and cool, neutral \vtwosig\ on the
    right ($-$700, $-$600, $-$500~\kms). The~minor-axis outflow
    reveals dusty filaments and unresolved gas motions of hundreds of
    \kms, consistent with a shocked outflow with layered clumps of
    gas. The~axes are in kpc. Reproduced by permission of the AAS from
    Figure 17 of reference \cite{2013ApJ...768...75R}.}
     \label{fig:f10565}
\end{figure}

Though they cannot yet match the numbers of galaxies observed in the
SDSS, spatially resolved spectroscopic surveys with optical IFSs
promise to significantly improve the detectability and
characterization of GWs. This is true both at ground-based resolution
and with enhanced spatial resolution from adaptive optics; the latter
is important for probing the nuclear regions of galaxies where GWs
emerge from their power sources. Studies of one or a few galaxies with
stellar GWs published in the current decade are numerous
\citep{2010ApJ...721..505R,2012ApJ...761..169F,2013ApJ...768...75R,2013ApJ...768..151V,2014MNRAS.444.3894H,2015MNRAS.448.2301M,2016MNRAS.461....6M,2017MNRAS.467.4951L,2017AA...604A.101C,2017AA...606A..95C}. These
detailed case studies highlight the ubiquity and complex, multiphase
nature (ionized, neutral, dusty) of these winds
(Figure~\ref{fig:f10565}) and the power of IFS for leveraging the
combination of spectral and spatial information to separate outflows
from their hosts.

Small surveys using instruments that target one galaxy at a time have
characterized GWs in galaxies with the highest star formation rates.
The authors of a study of the ionized gas in 27
LIRGs~\mbox{\citep{2014ApJ...781L..12R,2015ApJS..221...28R}} found
frequent shock ionization accompanied by high gas velocity
dispersions, a~conclusion supported by an earlier long-slit study of
ULIRGs~\citep{2012ApJ...757...86S}. The~line ratios are consistent
with models of slow shocks, correlate with $\sigma_\mathrm{gas}$, and
may be caused in part by GWs
\citep{2014ApJ...781L..12R,2015ApJS..221...28R}.

The authors of a larger study of the ionized and cool, neutral gas in
$\sim$50 LIRGs and ULIRGs~\citep{2014AA...568A..14A,2016AA...590A.125C} concur that high ionized
gas linewidths are shock-powered. To detect GWs, they spatially
integrated their spectra~\citep{2014AA...568A..14A,2016AA...590A.125C}
and/or fit the spatially resolved data~\citep{2016AA...590A.125C}. In
the former case, they corrected the velocity of each spatial position
for gravitational motions and then summed over the field of view. Fits
to the wind velocity vs. \sfr\ give best-fit slopes of 0.24 $\pm$ 0.05
(ionized gas), 0.15 $\pm$ 0.06 (integrated neutral gas), and
0.30 $\pm$ 0.05 (spatially resolved neutral gas). These are consistent
with the single-aperture fits discussed above (Section \ref{sec:uv}; Table
\ref{tab:uvfits}). $\eta$ is between 0.1 and 1 in the neutral and
ionized gas, with average values similar to those previously estimated
in LIRGs and ULIRGs
\citep{2005ApJS..160..115R,2013ApJ...768...75R}. The~study of
integrated spectra~\citep{2014AA...568A..14A} finds evidence that
$\eta$ decreases with increasing dynamical mass over the range
$10^{9.6}-10^{11.2}$~\msun, with a log-log slope of $-$0.43 that is
identical to that found in the UV over a much wider mass range
(\citep{2017MNRAS.469.4831C}; Section \ref{sec:uv};
Table~\ref{tab:uvfits}). (A steeper slope in this relationship at high
masses, as previously measured~\citep{2005ApJS..160..115R}, is
predicted by some simulations
\citep{2013MNRAS.436.1787L,2015MNRAS.452.1184M}.) Finally, these
authors fit a relationship between $\eta$ and \sigsfr\ with a log-log
slope of 0.17~\citep{2014AA...568A..14A}.

Two major (of order thousands of galaxies) IFS surveys have been
ongoing for several years: the SAMI
(Sydney-Australian-Astronomical-Observatory Multi-object
Integral-Field Spectrograph) Galaxy Survey~\citep{2015MNRAS.447.2857B}
and MaNGa (Mapping Nearby Galaxies at Apache Point Observatory)
\citep{2015ApJ...798....7B}. These surveys employ instruments that
target multiple galaxies at once, each with a single, small
IFS. Though limited in spatial resolution (of order 2\arcsec), the
ability of these surveys to integrate to high S/N, collect large
samples of galaxies across a wide parameter range, and separate the
outflow from the host galaxy using both spatial and spectral
information with IFS promises more accurate statements about the
ubiquity and property of stellar GWs over a wider range of galaxy
types.

The first result from these surveys relies on a sample of 40 edge-on
SAMI disk galaxies~\citep{2016MNRAS.457.1257H}. These authors use
minor-axis kinematic asymmetries to detect winds and find them down to
very low \sigsfr\ (10$^{-3}$~\smpy~kpc$^{-2}$). The~asymmetry
increases with \sigsfr\ and is associated with recent bursts of star
formation. The~connection of the asymmetry to GWs and its association
with high \sigsfr\ and high-temperature extraplanar gas is bolstered
by comparison to simulations~\citep{2018MNRAS.473..380T}.

\subsection{Molecular Gas and Dust} \label{sec:moldust}

The most rapid progress in the study of stellar GWs has come from
molecular gas and dust continuum measurements. The~direct dust
measurements have been made in the MIR/FIR using space-based
telescopes: {\it AKARI}, the {\it Spitzer Space Telescope}, and the
{\it Herschel Space Observatory}. The~molecular gas measurements have
been led by ground-based interferometers such as the Atacama Large
Millimeter/submillimeter Array (ALMA), SubMillimeter Array (SMA),
Institut de Radioastronomie Millim\'{e}trique (IRAM) Plateau de Bure
interferometer (PdBI), and Nobeyama Millimeter Array (NMA), though
{\it Herschel} has also played an important role. While sample sizes
are still small and most studies have focused on individual galaxies
in the starburst regime, results point to the ubiquity of dusty
molecular gas entrained in stellar GWs.

\begin{figure}[H]
  \centering
  \includegraphics[scale=0.72]{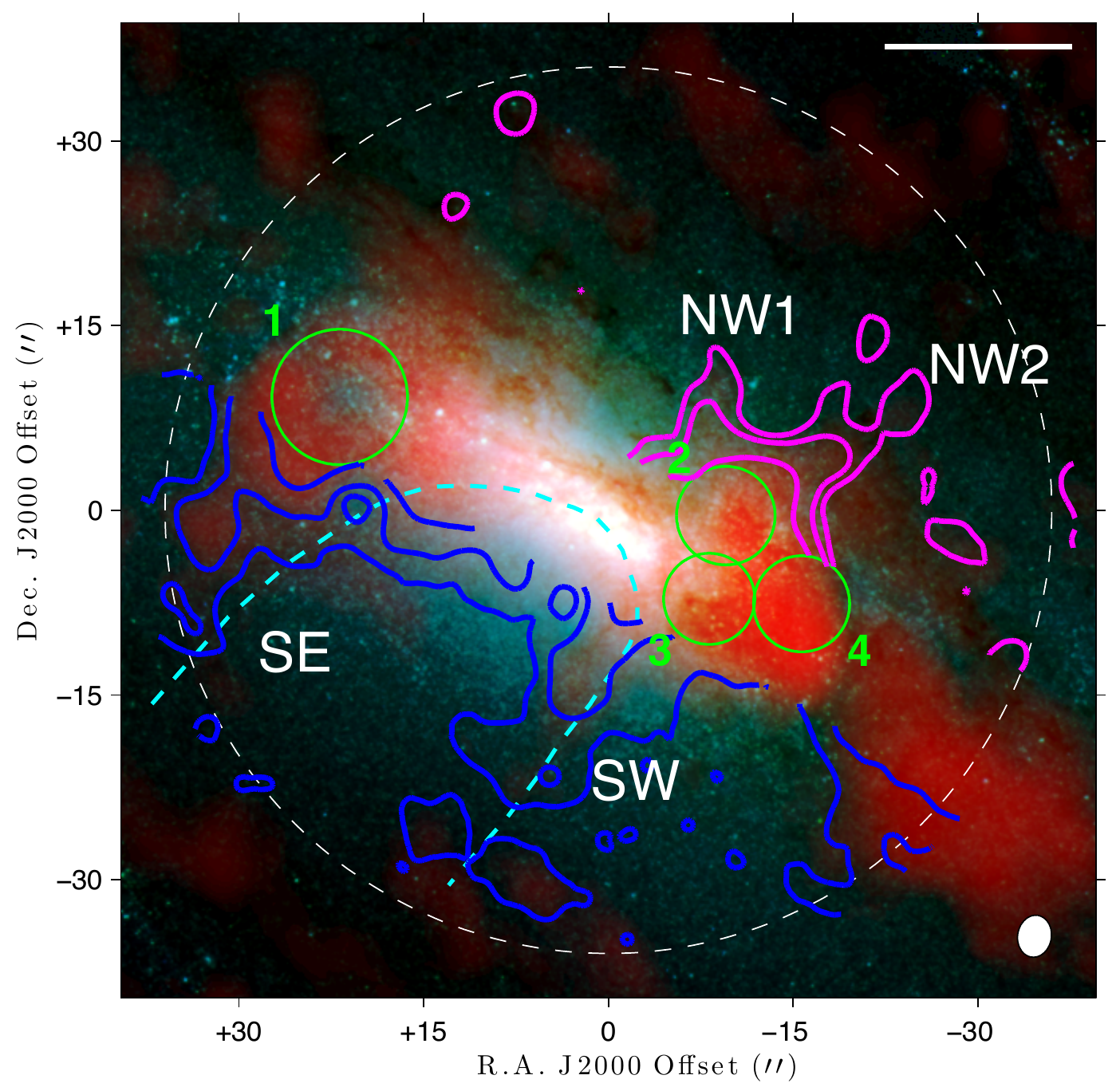}
  \caption{One of the nearest and best-resolved molecular gas outflows
    in NGC~253~\citep{2013Natur.499..450B}. The~background image is
    \hst\ NIR (blue$+$green) and ALMA CO (red). The~blue (magenta)
    contours show the approaching (receding) CO outflow. The~green
    circles outline expanding molecular shells. The~white bar is
    250~pc long. The~dashed cyan contour outlines the warm/hot ionized
    outflow that is interior to the molecular gas streamers and
    extends to larger scales (Section \ref{sec:moldust};
    ~\citep{2000AJ....120.2965S}). NGC~253 is one of only a few
    examples of the characteristic minor-axis filaments of a GW in
    molecular gas and their relationship to the disk and other gas
    phases. Reprinted by permission from reference
    \cite{2013Natur.499..450B}, \copyright2013.}
  \label{fig:ngc253}
\end{figure}

Recent interferometric measurements of CO transitions have uncovered
outflowing cold molecular gas in approximately 11 nearby LIRGs and
ULIRGs whose energy output is dominated by star formation
\citep{2009PASJ...61..237T,2014ApJ...797...90S,2014AA...562A..21C,2015AA...580A..35G,2016AA...594A..81P,2017ApJ...849...14S,2018ApJ...853L..28B,2018AA...609A..75F}.
Molecular outflows in four very nearby star-forming galaxies with
lower \sfr\ have also been studied in detail: NGC 3628, M82, NGC 253
(Figure~\ref{fig:ngc253}), and NGC 1808
\citep{2012ApJ...752...38T,2013Natur.499..450B,2015ApJ...814...83L,2016ApJ...823...68S}.
M82 had been the only previous starburst galaxy known to host a
molecular outflow~\citep{2001ApJ...563L..27G,2002ApJ...580L..21W}.

These molecular outflows are typically confined to the inner kpc of
the galaxy (in radius). They have modest outflow velocities in nearby
disks (tens to a few hundred \kms) but have a broader range of peak
velocities in the more luminous starbursting disks and mergers
(300--800~\kms). Similarly, the outflow rates appear to scale with
\sfr\ and/or \sigsfr. The~nearby, low-luminosity starbursts have
\mdot\ values of a few to a few tens of \smpy, while the LIRGs and
ULIRGs have a broader range of estimated \mdot: several to several
hundred \smpy\ (perhaps even $\sim$1000~\smpy\ in F17208$-$0014;~\citep{2015AA...580A..35G}). Mass outflow rates of these magnitudes
are similar to the star formation rates in these systems, and equal to
or larger than the mass outflow rates estimated in other gas
phases. The~larger mass outflow rates in the molecular phase are due
primarily to larger total outflowing gas masses.

Dense gas, as probed by characteristic molecules such as HCN and HCO$^+$,
is also a common feature of these outflows. It is present in the
extended outflows of M82, NGC 253, NGC 1808, and Arp 220, as~observed
in emission using interferometry
\citep{2014ApJ...797..134S,2014ApJ...780L..13K,2017ApJ...835..265W,2018ApJ...856...97S,2018ApJ...853L..28B}. Outflowing
dense molecular gas has also been observed in absorption in the inner
regions of the highly obscured binary Arp 220
\citep{2009ApJ...700L.104S,2015ApJ...800...25T,2016ApJ...833...41Z,2016AA...590A..25M}.

The ubiquity of starburst-driven molecular outflows was first
discovered using FIR OH and H$_2$O absorption lines with {\it
  Herschel}
\citep{2011ApJ...743...94R,2011ApJ...733L..16S,2013ApJ...775..127S,2013ApJ...776...27V,2017ApJ...836...11G}. The
outflow properties inferred from these spatially unresolved absorption
lines and detailed radiation transfer models are comparable to
measurements from molecular emission-line interferometry
\citep{2017ApJ...836...11G}: massive, high-velocity but compact flows.

The cold molecular and cool atomic gas share similar dynamics. Global
OH properties correlate with those measured from the 158~$\upmu$m
[\ion{C}{2}] emission line~\citep{2016ApJ...822...43J}, which traces
photo-dissociation regions (PDRs). The~158~$\upmu$m [\ion{C}{2}] line
may be a key technique for tracing outflows at high $z$ as it moves
into the submillimeter band. OH properties may also correlate with
\ion{Na}{1}~D atomic gas
measurements~\citep{2005ApJ...632..751R,2005ApJS..160..115R,2005ApJ...621..227M,2013ApJ...768...75R,2013ApJ...776...27V,2016ApJ...822...43J,2017ApJ...836...11G},
though there is significant scatter in this correlation. On the
spatially resolved level, FIR atomic fine structure lines are seen in
the GWs of M82~\citep{2013AA...549A.118C}, NGC
2146~\citep{2014ApJ...790...26K}, and NGC
253~\citep{2016AA...592L...3K}. In M82 and NGC 253 these atomic PDR
tracers correlate well with the molecular gas in terms of morphology
and dynamics~\citep{2013AA...549A.118C,2016AA...592L...3K}.

Direct observations of atomic \ion{H}{1} in outflows may become
possible with the next generation of sensitive radio arrays. New
Jansky Very Large Array \ion{H}{1} data on M82 are consistent with a
deceleration of the outflow as it moves upward into the halo
\citep{2018ApJ...856...61M}, suggestive of a galactic fountain
\citep{2015ApJ...814...83L}.

Warm molecular gas does not contribute significantly to the mass
budget of outflows, but the strength of its emission lines in the
accessible NIR band makes it a useful tracer of the extent and
physical state of molecular gas. Deep near-infrared (NIR) imaging and
spectroscopy shows that shocked, warm H$_2$ in the M82 outflow extends
several kpc from the disk along the minor axis~\mbox{\citep{2009ApJ...700L.149V,2015MNRAS.451.2640B}}. Outflowing warm
molecular gas has been detected in several nearby, starburst-dominated
ULIRGs using the 2.12~$\upmu$m H$_2$ 1-0 S(1) transition
\citep{2014AA...572A..40E,2015MNRAS.448.2301M,2017AA...607A.116E}. These
outflows are compact ($<$2 kpc), with~velocities of a few hundred
\kms, and the relationship between the cold and warm phases requires
further study. Indirect evidence from the global properties of a large
sample of ULIRGs implicates shocks in GWs as the origin of excess warm
H$_2$ in this population~\citep{2014MNRAS.439.2701H}.

Dust and molecular gas are abundant under similar physical
conditions. Observations of filamentary dust structures along disk
minor axes in some wind systems and the correlation of these dust
features with outflowing atomic gas columns are indirect evidence for
outflowing dust in stellar GWs (Figure \ref{fig:f10565}; e.g.,
\citep{1993AJ....105..486P,2010AJ....140..445C,2013ApJ...768...75R}).
Further evidence in the form of UV reflection nebula and polarized
line emission strongly suggests the presence of dusty winds in M82 and
NGC 253
\citep{2005ApJ...619L..99H,2011PASJ...63S.493Y,2014MNRAS.440..150H}.
Modeling of UV reflection in the M82 wind indicates smaller average
grain sizes in the M82 wind compared to the disk
\citep{2014MNRAS.440..150H,2015MNRAS.452.1412H}.

Observations of thermal emission from dust in galaxies where the disk
and outflow can be spatially separated strengthen the case that
stellar GWs are dusty in at least some cases. Studies detect
1--4$\times$10$^6$~\msun\ of dust in M82 (traced at 7--500~$\upmu$m)
extending far from the disk plane
\citep{2010AA...514A..14K,2010AA...518L..66R}, as well as significant
amounts of cold dust emplaced by tidal interaction with M81
\citep{2009AJ....137..517L}.

Roughly 10$^6$~\msun\ of warmer dust (traced at 70--160 $\upmu$m) is
also found in the outflows of NGC 253 and NGC 4136
\citep{2009ApJ...698L.125K,2015ApJ...804...46M} in structures that
correlate with ionized gas emission at other wavelengths. The~extended
cold dust in NGC 4631, however, may be of tidal origin
\citep{2015ApJ...804...46M}, as in the case of M82. In the dwarf
galaxy NGC 1569, 70--500 $\upmu$m imaging indicates a large reservoir (a
few times 10$^5$~\msun) of circumgalactic dust, perhaps deposited by
its starburst-driven GW~\citep{2018MNRAS.477..699M}. At shorter
wavelengths, GW features in NGC 1569 are seen in warm dust emission
\citep{2018MNRAS.477.3065S}. Contrary to these results, no~dust is
detected at 37$\upmu$m in the NGC 2146 wind
\citep{2014ApJ...790...26K}. FIR observations of dwarf galaxies with
outflows observed at other wavelengths paint a nuanced picture of
dusty outflows in systems with lower \mbox{\sfr\ \citep{2018MNRAS.477..699M}}; few systems show abundant circumgalactic
dust or strong correlations with multiwavelength signatures of GWs.

Finally, polycyclic aromatic hydrocarbons (PAHs) are prominent
features of GWs in those galaxies in which they have been detected. As
alternately large molecules or small dust grains, PAHs bridge the gap
between molecular gas and dust and are luminous features in rest-frame
3--20~$\upmu$m spectra. An~enormous PAH nebula extends from the M82
disk \citep{2006ApJ...642L.127E} and shows evidence for grain
shattering from PAH line ratios
\citep{2012AA...541A..10Y,2015MNRAS.451.2640B}. PAHs have been found
in GWs in other galaxies
\citep{2010AA...514A..15O,2013ApJ...774..126M}, and~the amount of
extraplanar PAH emission in galaxies may correlate with \sigsfr\
\citep{2013ApJ...774..126M}.

The warm and hot ionized gas traced by UV absorption lines, optical
emission lines, and X-rays are prevalent in GWs
(Sections \ref{sec:uv}--\,\ref{sec:ifs}). Quantifying the physical
relationship between the molecular and ionized gas phases thus appears
to be within reach. In a few nearby systems this connection has been
made, but there has been no attempt for most galaxies (probably
because the ability to detect the molecular phase of the outflow is
still relatively new). There is a need for sensitive, multiphase
studies at the highest spatial and spectral resolution to connect
these outflow components.

M82 and NGC 253 both have deep, high-spatial-resolution maps of warm
ionized gas (\ha), hot ionized gas (soft and hard X-rays), and
molecular gas. These two exemplars paint a picture of an innermost
hot, ionized wind fluid~\citep{2009ApJ...697.2030S} that entrains warm
and hot ionized material which are spatially correlated with each
other and surround the hot fluid, perhaps as a shell
\citep{1998ApJ...493..129S,2000AJ....120.2965S}. The~cold molecular
gas in turn envelops these ionized phases as it is entrained from the
disk
(Figure~\ref{fig:ngc253};~\citep{2013Natur.499..450B,2015ApJ...814...83L}). The~relationship
between the warm molecular and ionized phases, however, is likely more
complex. In examples of minor-axis outflows where both are spatially
resolved, there is coarse-grained spatial correlation (Arp 220
\citep{2015ApJ...810..149L,2018ApJ...853L..28B}, F08572$+$3915
\citep{2013ApJ...768...75R,2013ApJ...775L..15R}, and M82
\citep{2009ApJ...700L.149V}), but in other cases there is not (NGC
4945;~\citep{2000AA...357...24M}). At very high spatial resolution,
apparent correlation may break down \citep{2009ApJ...700L.149V}. Note
that the outflows in some of these galaxies may be AGN-driven, but
they are included here because the sample size of galaxies with
resolved molecular and ionized outflows is unfortunately~small.

\section{The Nearest Galactic Wind}

The unexpected recent discovery of the so-called Fermi bubbles--large,
diffuse $\gamma$-ray structures that form a bipolar shape above and
below the Galactic plane--has revived interest in characterizing the
GW in the Milky Way
\citep{2010ApJ...724.1044S,2014ApJ...793...64A}. Previous data
provided strong evidence of its existence (see references in
\citep{2005ARAA..43..769V}), but the picture-perfect morphology of the
Fermi bubbles make it an almost inarguable fact.

This review is concerned with observations of starburst-driven GWs,
and the Milky Way's GW may be driven by star formation
(e.g.,~\citep{2015ApJ...808..107C}). It also may be powered by the
Galactic nuclear black hole during a previous accretion episode (see,
e.g., the recent review of relevant data and models in
\citep{2018Galax...6...29Y}). However, for completeness we note some
recent observations, since the Milky Way is an excellent laboratory
for studying a GW at high sensitivity and spatial resolution in what
is a ``typical'' galaxy in the Local Universe. Since its discovery,
the Fermi bubbles have notably been found to also contain a magnetized
radio plasma~\citep{2013Natur.493...66C} and have been connected to
the previously known microwave ``haze'' that was re-observed with {\it
  Planck}~\citep{2013AA...554A.139P}. The~bubbles also appear to host
neutral gas clouds moving up to several hundred \kms\
\citep{2013ApJ...770L...4M,2016ApJ...826..215L,2018ApJ...855...33D},
as well as higher-ionization species observed in absorption
\citep{2014ApJ...785L..24F,2017ApJ...834..191B,2017ApJS..232...25S,2018ApJ...860...98K}. These
neutral and ionized clouds may lie along filaments swept up by the
bubble along its edges, though their contribution to the structure and
mass/energy budget of the outflow are not yet clear.

A recent claim has also been made for a GW in a satellite of the Milky
Way. This potential outflow in the Large Magellanic Cloud was
extrapolated from absorption-line measurements of a single line of
sight through its disk~\citep{2016ApJ...817...91B}. The~LMC contains
30 Doradus, which is undergoing a significant star formation episode.

\section{Winds Driven by Star Formation at High redshift}

The presence and ubiquity of stellar GWs outside of the local universe
was first evident in blueshifted rest-frame UV absorption lines and
complex Ly$\alpha$ profiles in Lyman-break galaxies (LBGs)
(e.g.,~\citep{2000ApJ...528...96P,2001ApJ...554..981P,2010ApJ...717..289S}).
Information on stellar GWs at high $z$ comes predominantly from
single-aperture, down-the-barrel spectroscopy of rest-frame UV
lines. (``Down-the-barrel'' refers to sightlines toward the galaxy
itself.) However, newer instruments and techniques have in the past
decade opened other avenues to study high $z$ winds. These include
rest-frame optical measurements of emission lines with wide-field,
NIR, multi-object spectrographs; multiplexed or wide-field IFS
instruments such as the K-band Multi-Object Spectrograph (KMOS) and the
Multi-Unit Spectroscopic Explorer (MUSE) that enable multi-object
and/or spatially resolved measurements; NIR, adaptive optics IFS to
achieve high spatial resolution; FIR and submillimeter observations
that probe molecular and atomic gas transitions; and
transverse-sightline spectroscopic surveys. These new techniques have
established the ubiquity and properties of stellar GWs in new galaxy
populations and constrained the redshift evolution of their bulk
properties.

Deep spectroscopy of strong rest-frame optical emission lines (mainly
\ha\ and \nt) at $z\sim2$ reveals broad wings that arise primarily
from bright star-forming regions
\citep{2011ApJ...733..101G,2011AA...534L...4L,2012ApJ...752..111N,2012ApJ...761...43N}. These
broad wings, which appear to extend over several kpc and strengthen
with increasing \sigsfr, have been interpreted as evidence of stellar
feedback. Similar wings are found in massive, compact star-forming
galaxies over a wider wavelength range~\citep{2018ApJ...855...97W}.

\begin{figure}[h]
  \centering
  \includegraphics[width=\textwidth]{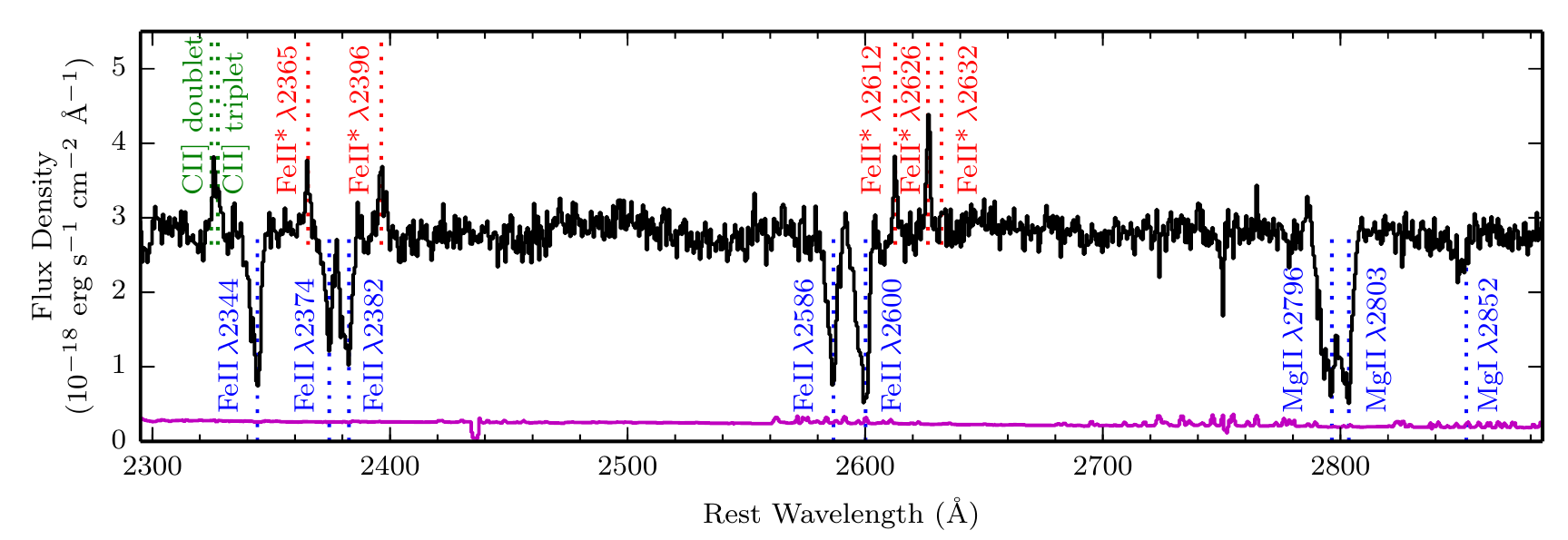}
  \caption{Absorption and emission lines in the outflow of a $z = 1.3$
    star-forming galaxy~\citep{2017AA...605A.118F}. The~resonant
    absorption lines (labeled in blue) are blueshifted and trace the
    approaching near side of the outflow. The~non-resonant iron
    emission lines (labeled in red), which are spatially extended
    along the minor axis, trace the bulk of the outflow. These
    resonant and non-resonant lines are powerful probes of the
    presence and properties of high-redshift outflows. Reproduced with
    permission from Figure 1 of reference \cite{2017AA...605A.118F},
    \copyright ESO.}
     \label{fig:finley}
\end{figure}

Larger spectroscopic surveys (up to $\sim$500 galaxies) of
star-forming galaxies at $z = 0.3-2$ use the low-ionization metal
lines \ion{Mg}{1}, \ion{Mg}{2}, and \ion{Fe}{2} to probe GWs
\citep{2009ApJ...692..187W,2010ApJ...719.1503R,2011MNRAS.418.1071B,2012ApJ...758..135K,2012ApJ...759...26E,2012ApJ...760..127M,2013MNRAS.433..194B,2014ApJ...794..130B,2014ApJ...794..156R,2017AA...608A...7F}. Resonant
transitions from outflows in absorption (blueshifted) and emission
(redshifted), as well as corresponding non-resonant transitions,
constrain basic outflow properties such as velocity and ionization
state and, potentially, more~complex structural parameters
(Figure~\ref{fig:finley}; \citep{2011ApJ...734...24P}). These
high-$z$, low-ionization outflows are broadly consistent with those
found at low $z$: they have modest velocities (up to a few
hundred~\kms\ on average); their properties (velocity and equivalent
width) correlate with \sfr, \sigsfr, and \mstar; their detection rates
in absorption average a few tens of percent, indicative primarily of
the wind geometry (a high frequency of occurrence of non-spherical
winds); they have estimated mass-loss rates of order the star
formation rate, though with considerable uncertainty; and they are
preferentially found in face-on galaxies (or~their properties are more
extreme in face-on galaxies), consistent with minor-axis
flows. The~velocities of these winds may increase with increasing $z$
for galaxies of given \sfr, possibly due to increasing star formation
rate surface density
\citep{2012ApJ...759...26E,2015ApJ...800...21W,2017ApJ...850...51S,2018ApJ...860...75D},
and with increasing ionization
potential~\citep{2016ApJ...829...64D,2018ApJ...860...75D,2018MNRAS.474.1688C}. Extended,
scattered emission (both resonant and non-resonant) from
low-ionization species has been probed in detail in a handful of
systems
\citep{2011ApJ...728...55R,2013ApJ...774...50K,2016MNRAS.458.1891B,2017AA...605A.118F}
and is now recognized as a common feature of star-forming galaxies at
these epochs
\citep{2011ApJ...743...46C,2013ApJ...770...41M,2017AA...608A...7F}. Pure
resonant emission is seen at low \sfr, transitioning to P-Cygni or
pure-absorption resonant profiles plus non-resonant emission at high
\sfr, indicative of an increasing signature of outflowing gas
\citep{2017AA...608A...7F}. In a handful of gravitationally lensed
systems, absorption-line analyses of more ions indicate $\alpha$
enhancement and more robustly constrain mass-loss rates
\citep{2002ApJ...569..742P,2018ApJ...863..191J}.

At moderate redshifts ($z\sim0.7-0.8$), the presence of low-velocity
GWs in poststarburst galaxies points to the possibility that these
winds help to quench star formation~\citep{2011ApJ...743...46C}. Very
high-velocity winds (1000~\kms) found in bluer, rarer post-starbursts
\citep{2007ApJ...663L..77T} appear to be driven by very compact
starbursts rather than AGN activity
\citep{2012ApJ...755L..26D,2014MNRAS.441.3417S}.

The down-the-barrel technique, which probes low-ionization outflows
that emerge along the line-of-sight toward the galaxy disk, is
complemented by transverse sightlines through galaxy halos toward
background quasars or galaxies. LBGs at $z = 2-3$ host both low- and
high-ionization gas out to radii $\sim$100~kpc; because this gas is
outflowing over a large solid angle (as inferred from down-the-barrel
observations), it is plausibly doing so at large radii
\citep{2010ApJ...717..289S}. In other samples, \ion{Mg}{2} and
\ion{O}{6} absorbers show a preference for alignment with galaxy major
or minor axes, and the minor-axis gas is preferentially found in blue
galaxies. This absorber alignment is suggestive of major-axis inflow
and minor-axis outflow, and the connection to blue galaxies points to
star formation as the power source
\citep{2012ApJ...760L...7K,2015ApJ...815...22K}. Simple geometric
outflow models of halo absorbers yield outflow properties that are
consistent with those seen at low $z$
\citep{2012MNRAS.426..801B,2015ApJ...804...83S,2015ApJ...811..132M}.

The prominence of \lya\ in the spectra of high $z$ galaxies makes it a
tempting target for parameterizing outflows
(e.g.,~\citep{2015Natur.523..169E}). However, as mentioned above
(Section \ref{sec:uv}), radiation transfer effects make it an
ambiguous indicator. Redshifted Ly$\alpha$ does typically accompany
blueshifted low-ionization lines
(e.g.,~\citep{2010ApJ...717..289S}). Star forming galaxies also show
an increase in the velocity of Ly$\alpha$ as \sfr\ and Ly$\alpha$
equivalent width increase
\citep{2013ApJ...765...70H,2014ApJ...788...74S,2014ApJ...795...33E}. However,
whether this is due solely to changing outflow properties or instead
to an increase in gas near the systemic redshift is unclear
\citep{2014ApJ...795...33E}.

Finally, a handful of molecular gas detections of outflows at
moderate-to-high $z$ are emerging. CO has been imaged in two
high-velocity, apparently stellar GWs in post-starbursts at $z\sim0.7$
\citep{2014Natur.516...68G,2018ApJ...864L...1G}. A~long {\it Herschel}
integration allowed detection of an OH outflow in absorption in a
$z=2.3$ ULIRG~\citep{2014MNRAS.442.1877G}. A~serendipitous ALMA
discovery of extremely broad CH$^+$ in several $z\sim2.5$ ULIRGs
points to turbulent outflows~\citep{2017Natur.548..430F}. Tentative
detections of broad, faint [\ion{C}{2}] line wings at $z = 5.5$ hint
at the possibility of stellar GWs in modestly star-forming galaxies at
this epoch~\citep{2018MNRAS.473.1909G}. Finally, an OH outflow exists
in a gravitationally lensed, dusty galaxy at $z = 5.3$
\citep{2018Sci...361.1016S}.

\section{Summary}

We can say with reasonable certainty that GWs driven by
energy from stellar processes are a common feature of galaxies with
moderate-to-high star formation rates and/or surface densities out to
$z\sim2-3$. Stacking analyses of large rest-frame UV and optical
spectroscopic surveys have established that the average star-forming
galaxy has an ionized and/or neutral wind whose velocity scales with
star formation rate, stellar mass, and possibly \sigsfr. At low
redshift, these winds are most prominent in starburst galaxies that
lie above the galaxy main sequence; at higher redshift, where galaxies
on the main sequence have higher \sfr, the situation may be
different. Hints exist that GWs are common but simply hard to detect
even in galaxies with low \sigsfr\ (for instance, the~low-surface-brightness Milky Way GW).

The correlations between outflow and galaxy properties found in some
of the first large surveys of stellar GWs have been verified and
refined by larger and more diverse samples and different gas
probes. Besides serving merely as input to parameterizations of
outflows in numerical simulations, measurements of GWs can now be
compared to the predicted properties of GWs from simulations that
better implement the physics of stellar feedback.

Detailed, multiwavelength studies of star-forming galaxies continue to
reveal new layers of GWs. Most notable is that stellar GWs entrain
large quantities of molecular gas, including dense clumps, and loft
dust and soot (PAHs) far above the galactic disk. The~promising
technique of combining resonant-line absorption and emission with
non-resonant re-emission channels has been successfully used to detect
winds at high $z$ and may prove a powerful probe of GW structure and
extent when widely deployed. Observations of a wider range of galaxies
besides the usual suspects (e.g., M82) with 3D imaging spectroscopy
shows that a complex, multiphase structure of filaments of dusty
ionized and neutral gas collimated along the minor axis is a common
feature of GWs. Transverse-sightline spectroscopy and correlations
with galaxy inclination at a variety of redshifts bolster this
picture. Finally, increasingly in-depth studies of local galaxies with
extended Ly$\alpha$ and LyC may eventually put meaningful constraints
on how outflows contribute to reionization and help interpret high-$z$
observations of Ly$\alpha$.

Future progress will occur on a variety of fronts. At low $z$, a new
generation of ongoing multi-object IFS surveys (SAMI Galaxy Survey,
MaNGA) will soon produce results on thousands of nearby
galaxies. Future, much larger IFS surveys are being planned (using,
e.g., Hector;~\citep{2016SPIE.9908E..1FB}). Sensitive, wide-field IFS
instruments on large telescopes (such as MUSE and KCWI, the Keck
Cosmic Web Imager) will probe the full extent of GWs in nearby,
well-resolved targets and enable efficient, spatially resolved
characterization of many galaxies at once in high-$z$ deep
fields. Next generation multi-object spectroscopy surveys (e.g., the
Dark Energy Spectroscopic Instrument, or DESI, Survey and 4MOST, the
4-metre Multi-Object Spectroscopic Telescope) will increase the
fidelity of stacking analyses over a wider range of redshift and
galaxy properties.

Continued measurements with ALMA, particularly at high spatial
resolution, will provide more detailed understanding of the structure
and chemistry of molecular gas in outflows. ALMA will also certainly
expand on its currently short list of detections of high-$z$ stellar
GWs. The~high resolution and sensitivity of the {\it James Webb Space
  Telescope} ({\it JWST}) in the MIR will undoubtedly produce useful
measurements of molecular gas and dust in stellar GWs, as
well. However, the spatial resolution and sensitivity of {\it JWST} is
likely to provide the most dramatic advances in measuring the
properties of outflows in high-$z$ galaxies by characterizing them in
individual main-sequence galaxies at $z\sim2-3$ and detecting them at
very high $z$, where their impact could be especially significant but
where measurements currently do not exist.

Finally, we note two areas of study that have seen little recent
progress, but whose prospects should eventually rise. Measurements of
the radio-emitting plasma in GWs are very rare except for a
few recent detections
\citep{2010ApJ...712..536K,2017MNRAS.471.2438L,2018AA...610L..18R}. The
next generation of wide-field radio arrays may make this a growth
area. The~field of X-rays studies of GWs has also lain fallow, with a
few exceptions
(e.g.,~\citep{2014ApJ...793...73O,2018MNRAS.477.3164M}). The~hottest
gas phase of GWs, which may drive the outflows in starburst galaxies,
has proven extremely difficult to detect except in the nearest cases
\citep{2009ApJ...697.2030S}. More sensitive X-ray telescopes in the
coming two decades will eventually lead to a better characterization
of this pivotal component.
\vspace{6pt}


\acknowledgments{The author thanks the referees for their feedback.}

\funding{This research received no external funding. The~author is
  supported by the J. Lester Crain chair at Rhodes College.}

\conflictsofinterest{The author declares no conflict of interest.}

\abbreviations{The following abbreviations are used in this manuscript:\\
\noindent
\begin{tabular}{@{}ll}
  ALMA & Atacama Large Millimeter/submillimeter Array\\
  COS & Cosmic Origins Spectrograph\\
  FIR & far-infrared\\
  GW & galactic wind\\
  IFS & integral field spectrograph\\
  LBG & Lyman-break galaxy\\
  LIRG & luminous infrared galaxy\\
  MIR & mid-infrared\\
  NIR & near-infrared\\
  SFR & star formation rate\\
  sSFR & specific star formation rate\\
  ULIRG & ultraluminous infrared galaxy
\end{tabular}}

\reftitle{References}

\end{document}